\documentclass[
  superscriptaddress,
  amsmath,amssymb,
  aps,
  reprint,
  prb,
]{revtex4-1}

\usepackage{graphicx}  
\usepackage{siunitx}
\usepackage{caption}
\usepackage{subcaption} 
\usepackage{dcolumn}
\usepackage{bm}
\usepackage[normalem]{ulem}
\usepackage[colorlinks=true, citecolor=blue, linkcolor=blue, urlcolor=blue]{hyperref}
\usepackage{amsmath}
\usepackage[capitalize]{cleveref}
\usepackage{comment}

\newcommand{\tc}{$T_{\text{c}}\,$}
\newcommand{\rs}{$R_{\text{s}}\,$}

\newcommand{\idet}{$I_{\text{det}}\,$}

\newcommand{\irel}{$I_{\text{rel}}\,$}
\newcommand{\idep}{$I_{\text{dep}}\,$}
\newcommand{\iret}{$I_{\text{ret}}\,$}
\newcommand{\ibias}{$I_{\text{bias}}\,$}

\newcommand{\mosi}[2]{Mo$_{\text{#1}}$Si$_{\text{#2}}$}
\newcommand{\sub}[2]{${#1}_{\text{#2}}$}

\begin{document}

\title{Impact of Stoichiometry of MoSi Thin Films for Enhanced Sensitivity of Superconducting Nanowire Single-Photon Detectors}

\date{\today}

\author{Stefanie Grotowski}
\email{stefanie.grotowski@tum.de}
\affiliation{Walter Schottky Institut, Technische Universität München, 85748 Garching, Germany}
\affiliation{TUM School of Natural Sciences, Technische Universität München, 85748 Garching, Germany}

\author{Damjan Pecijareski}
\affiliation{Walter Schottky Institut, Technische Universität München, 85748 Garching, Germany}
\affiliation{TUM School of Natural Sciences, Technische Universität München, 85748 Garching, Germany}

\author{Hadrien Le Petit Delacour}
\affiliation{Walter Schottky Institut, Technische Universität München, 85748 Garching, Germany}
\affiliation{TUM School of Natural Sciences, Technische Universität München, 85748 Garching, Germany}

\author{Lucio Zugliani}
\affiliation{Walter Schottky Institut, Technische Universität München, 85748 Garching, Germany}
\affiliation{TUM School of Computation, Information and Technology, Technische Universität München, 80333 München, Germany}

\author{Fabian Wietschorke}
\affiliation{Walter Schottky Institut, Technische Universität München, 85748 Garching, Germany}
\affiliation{TUM School of Computation, Information and Technology, Technische Universität München, 80333 München, Germany}

\author{Christian Schmid}
\affiliation{Walter Schottky Institut, Technische Universität München, 85748 Garching, Germany}
\affiliation{TUM School of Computation, Information and Technology, Technische Universität München, 80333 München, Germany}

\author{Stefan Strohauer}
\affiliation{Walter Schottky Institut, Technische Universität München, 85748 Garching, Germany}
\affiliation{TUM School of Natural Sciences, Technische Universität München, 85748 Garching, Germany}

\author{Matthias Althammer}
\affiliation{Walther-Meißner-Institut, Bayerische Akademie der Wissenschaften, 85748 Garching, Germany}

\author{Rudolf Gross}
\affiliation{Walther-Meißner-Institut, Bayerische Akademie der Wissenschaften, 85748 Garching, Germany}
\affiliation{Munich Center of Quantum Science and Technology (MCQST), 80799 München, Germany}

\author{Kai Müller}
\affiliation{Walter Schottky Institut, Technische Universität München, 85748 Garching, Germany}
\affiliation{TUM School of Computation, Information and Technology, Technische Universität München, 80333 München, Germany}
\affiliation{Munich Center of Quantum Science and Technology (MCQST), 80799 München, Germany}

\author{Jonathan J. Finley}
\affiliation{Walter Schottky Institut, Technische Universität München, 85748 Garching, Germany}
\affiliation{TUM School of Natural Sciences, Technische Universität München, 85748 Garching, Germany}
\affiliation{Munich Center of Quantum Science and Technology (MCQST), 80799 München, Germany}

\begin{abstract}
    We report on the impact of the stoichiometry of superconducting MoSi thin films on the performance of superconducting nanowire single-photon detectors (SNSPDs).
    Specifically, we investigate the relation between the film parameters critical temperature \tc, sheet resistance \rs and superconductor thickness $d$ and observe a universal scaling behavior. 
    To benchmark the performance of SNSPDs fabricated from films having different stoichiometry, we measure the bias dependent count rate curves, while the detector is illuminated with wavelengths between \SI{780}{\nano \meter} and \SI{1550}{\nano \meter}. 
    The detector performance as a function photon energy for different nanowire widths reveals a linear relation between the detection current and the photon energy.
    Furthermore, we determine the interfacial thermal boundary conductance $\beta$ between the superconducting thin film and the substrate, by measuring the return current of the SNSPD and find an increase of $\beta$ with increasing Mo concentration. 
    The highest sensitivity amongst all compared devices is achieved for \mosi{0.53}{0.47}, with low \tc (\SI{4.1}{\kelvin}) and high \rs (\SI{397}{\ohm / sq}) at a film thickness of \SI{5.4}{\nano \meter}. 
\end{abstract}

\maketitle

\section{Introduction}
\vspace{-10pt}
Superconducting Nanowire Single-Photon Detectors (SNSPDs) \cite{Goltsman2001} are a crucial building block for the realization of photonic quantum technologies. 
Their ability to detect single-photons with high efficiency \cite{Reddy2020}, low dark count rate \cite{Shibata2015} and high timing resolution \cite{Korzh2020} makes them one of the most performant single-photon detectors. 
Silicide based superconductors have been used as a superconducting material for several years, especially for applications in the NIR and IR regime, due to their smaller superconducting energy gap compared to more widely used materials such as NbN and NbTiN. 
In addition, the amorphous structure of silicide-based materials promises higher fabrication yield \cite{Allman2015}.
Thereby, MoSi is one of the major material platforms used for the fabrication of SNSPDs. 
Although the superconducting and electrical transport properties of this material have been studied in various publications, their translation onto SNSPD device performance, especially over this range of stoichiometries from \mosi{0.53}{0.47} to \mosi{0.74}{0.26}, has not been investigated yet. 
Therefore, our aim is to systematically investigate the dependence of sensitivity of SNSPDs on the MoSi stoichiometry. 
We begin by giving a brief overview of the metrics of the superconducting thin films, as they form the basis of our devices. 
We then continue to present investigations on the sensitivity, based on the hotspot model established by Semenov \textit{et al.} \cite{Semenov2005}.
Regarding the characterization of SNSPD performance we focus on the analysis of the bias current dependent count rate curves.
Compared to previously used techniques to quantify the sensitivity, where the cutoff wavelength at a given relative bias current is identified, our approach focuses on determining the detection current, at a given wavelength. 
This method offers two major advantages: Firstly, it obviates the need for expensive tunable laser sources.
Secondly, it is particularly advantageous for characterizing devices where the cutoff wavelengths are beyond the telecom C-band, where complex optical instrumentations are necessary and where laser illumination begins to overlap spectrally with blackbody radiation. 
Importantly, we find that the device with lowest Mo concentration provides highest sensitivity.
Furthermore, we calculate the thermal boundary conductance $\beta$ from measurements of the return current to investigate the thermal coupling between the superconductor and the substrate. 
We compare our values with calculations, based on film parameters from literature and we show that the predicted increase of $\beta$ with Mo concentration in the framework of the diffusive mismatch model holds true. 
Finally, we fit our results of the detection current to find the parameter representing the conversion efficiency of photon energy to a detection event.
The highest conversion efficiency is found for the highest Mo concentration.

\section{Methods}
\vspace{-10pt}
The samples are deposited by magnetron co-sputtering in an ultra-high vacuum chamber at pressures of around \SI{1.4e-8}{\milli \bar}. 
Both molybdenum (Mo) and silicon (Si) targets have a diameter of 2 inch and are controlled independently. 
Si is deposited with a constant RF power of \SI{75}{\watt}. 
A DC power between 24-\SI{57}{\watt} is used for the Mo target, while the working pressure during deposition is \SI{3.1e-3}{\milli \bar}. 
We deposit thicknesses between \SI{4.9}{\nano \meter} and \SI{5.4}{\nano \meter} on a Si substrate with a \SI{130}{\nano \meter} thick thermally oxidized SiO$_2$ on the deposition surface. 
To prevent oxidation of the thin film it is covered by a \SI{4}{\nano \meter} amorhous Si-layer. 
The thickness of the layers is controlled via a quartz crystal measuring the deposition rate and the resulting thickness is precisely determined by x-ray reflectivity (XRR) measurements. 
The thickness uncertainty is given by the resolution of the XRR machine, which is \SI{0.4}{\nano \meter}. 
Furthermore, we determine the stoichiometry using X-ray photoelectron (XPS) spectroscopy measurements. 
We determine the sheet resistance \rs from a 4 point measurement with an uncertainty of 2\%. 
\tc is determined by measuring the temperature dependence of the resistance via a 4 point probe from room temperature to $<$ \SI{3}{\kelvin}. 
The transition temperature \tc is defined as the temperature where the resistance of the film is half of the normal conducting resistance at \SI{20}{\kelvin}. 
The standard deviation of \tc is determined by multiple temperature sweeps. 
\tc of a nanowire is determined via a 2 point measurement. 
The quasiparticle diffusivity $D$ is extracted from measurements of the superconducting transition at different applied magnetic fields, varying from \SI{-0.1}{\tesla} to \SI{1.1}{\tesla} as described by Strohauer \textit{et al.} \cite{Strohauer2023}. 

The SNSPD fabrication starts with patterning the nanowire structures into resist \textit{AR-N 7520.073} via electron beam lithography. 
The pattern is then transferred to the superconducting film via reactive ion etching in a CF$_4$ environment. 
Electrical contacts are established using optical lithography and evaporation of Ti/Au contact pads. 
The superconducting nanowire width of the devices is varied between \SI{65}{\nano \meter} to \SI{190}{\nano \meter} for all thin films. 
Subsequently, we characterize the samples inside an ADR cryostat (\textit{Kiutra GmbH}) at \SI{1}{\kelvin}. 
The devices are illuminated with calibrated continuous wave lasers at wavelengths of \SI{780}{\nano \meter}, \SI{930}{\nano \meter}, \SI{1310}{\nano \meter} or \SI{1550}{\nano \meter}, respectively. 
To avoid latching, a \SI{25}{\ohm} resistor is connected in parallel to the detector. 
The output signal is amplified at \SI{3}{\kelvin} (\textit{Cosmic Microwave Technology, CITLF3}) and at room temperature (\textit{RF Bay, LNA2000}), and then analyzed using a pulse counting device. \\

\section{Results}
\begin{figure}[h]
    \centering
    \includegraphics[width=\columnwidth]{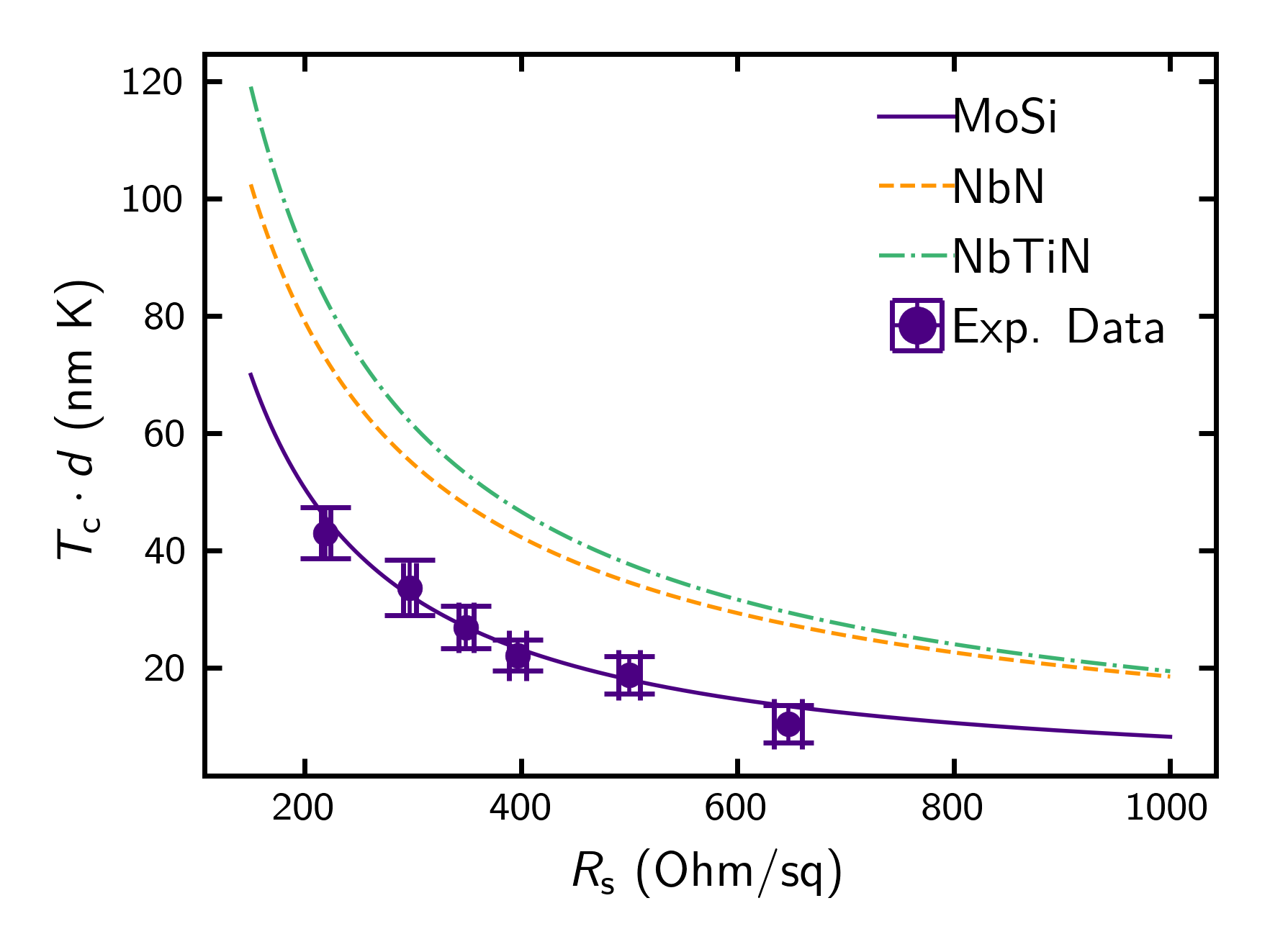}
    \caption{Universal scaling law for MoSi thin films used in this experiment. The purple line is a fit to our data according to \ref{eq:usl}. The orange and green curve are literature data.}
    \label{fig:universal}
\end{figure}

\vspace{-10pt}
We start with an overview of the superconducting films.  
The product of \tc and film thickness $d$ is presented as a function of sheet resistance \rs in Figure~\ref{fig:universal} as purple data points. 
The observed behavior can be described with a universal scaling law \cite{Ivry2014} between the film parameters \tc, \rs and $d$
\begin{equation}
    T_{\text{c}}[\mathrm{K}] \cdot d [\mathrm{nm}] = A \cdot R_{\text{s}}[\Omega / \mathrm{sq}]^{-B}. 
    \label{eq:usl}
\end{equation}
Given that the variables are provided in K, $\Omega$/sq and nm, respectively, the parameters $A$ and $B$ are dimensionless and material dependent. 
A fit to the data (purple line) yields $A\,=\,20214$ and $B\,=\,1.13$.
Zhang \textit{et al.} \cite{Zhang2021} applied the universal scaling law to \mosi{x}{1-x} thin films with $x$ ranging from 0.26 to 0.83 including different thicknesses. 
Their analysis indicated that even with varying stoichiometry, the system can be characterized by a single, universal pair of fitting parameters. 
The parameter $B$ is larger than 1, which indicates an amorphous film structure \cite{Ivry2014} and is in agreement with values reported in other studies \cite{Banerjee2017, Zhang2021}. 
For comparison, we also present in Figure~\ref{fig:universal} the scaling behavior for NbN \cite{Ivry2014} (orange curve) and NbTiN \cite{Strohauer2023} (green curve).
The achievable combination of \tc and \rs for MoSi is lower than those for NbN and NbTiN, respectively. This enables higher sensitivity for lower photon energies using MoSi, one of major the reasons, why silicide based materials are most promising for SNSPDs in the mid- to infrared regime \cite{Verma2021}.

We now turn to the characterization of SNSPDs fabricated from three films having different Mo concentrations ranging from 53\% to 74\%. 
The film dependent detector parameters are summarized in Table~\ref{tab:films}. 
Since \tc is known to decrease with decreasing width of nanowire structures \cite{Banerjee2018, Chen2022}, we measured the critical temperature of our films for different nanowire widths.
The superconducting transition was measured at current densities below \SI{2.72e5}{\ampere \per \centi \meter ^2}. 
Our results show, as an example, a reduction of \tc from \SI{6.55}{\kelvin} to \SI{6.15}{\kelvin} for a thin film compared to a nanowire. 
However, the \tc variation within our nanowire width variation was smaller than the uncertainty stemming from the measurement. 
In the table we state the mean \tc of the nanowires, as it will be used in all subsequent calculations. 

\begin{table}[h]
    \centering
    \begin{tabular}{|c|c|c|c|c|}
         \hline
        Film & Mo concentration & $d$ (nm) & \rs ($\Omega /$sq) & \tc (K) \\ \hline
        A & 0.53 & 5.4 & 397 & 4.11 \\ \hline
        B & 0.61 & 4.9 & 349 & 5.25 \\ \hline
        C & 0.74 & 5.1 & 297 & 6.15 \\ \hline
    \end{tabular}
    \caption{Summary of thin film parameters used for SNSPD fabrication. }
    \label{tab:films}
\end{table}

In the following, we first provide information about the models used to analyze the dependence of the count rate on the bias current for identical SNSPD geometries and different stoichiometries. 
In the framework of the diffusive hotspot model the cutoff wavelength of a SNSPD is related to the minimum energy required to break the Cooper pairs and to generate a hotspot \cite{Semenov2005}. 
The spectral cutoff can be derived within the hotspot model by considering the diffusion of quasiparticles generated after photon absorption. 
The absorbed photon breaks up Cooper pairs and creates a cloud of quasiparticles that diffuses within the superconducting nanowire. 
A resistive state is triggered when the resulting reduction of the superconducting pair density causes the local current density to exceed the critical value. 
The cutoff wavelength is obtained by relating the number of quasiparticles produced by the absorbed photon with the minimum number required to suppress superconductivity within a segment of the wire of length comparable to the coherence length. This yields:
\begin{equation}
    \lambda_{\text{c}} = \frac{\xi R_{\text{s}} e^2 h c}{\Delta^2 w} \cdot \sqrt{\frac{D}{\pi \tau_{\text{th}}}} \left( 1- \frac{I_{\text{bias}}}{I_{\text{dep}}}\right)^{-1},
    \label{eq:cutoff}
\end{equation}
where $D$ is the diffusivity, $\Delta$ the superconducting energy gap, $\xi$ the fraction of absorbed energy deposited in the hotspot, and $\tau_{\text{th}}$ the electron thermalization time. \ibias is the bias current of the detector and \idep the depairing current.
Equation~\eqref{eq:cutoff} can be rearranged to determine the minimum bias current required for a photon of wavelength $\lambda$ to generate a hotspot:

\begin{equation}
I_{\mathrm{min}}(\lambda) = I_{\mathrm{dep}}
\left[1-\frac{1}{\lambda}\left(\frac{\xi R_{\text{s}} e^2 h c}{\Delta^2 w} \cdot \sqrt{\frac{D}{\pi \tau_{\text{th}}}}\right)
\right] 
\label{eq:detection_current}
\end{equation}
We define the minimum current $I_{\mathrm{min}}(\lambda)$ normalized by \idep as the normalized detection current \idet 

\begin{equation}
    I_{\mathrm{det}}(\lambda) = \frac{I_{\mathrm{min}}(\lambda)}{I_{\mathrm{dep}}}.
\end{equation}
The depairing current density is calculated using the Ginzburg-Landau expression with the dirty correction of Kupryanov and Lukichev \cite{Kupriyanov1980}:

\begin{equation}
J_{\mathrm{dep}}(T) = J_{\mathrm{dep}}(0) \, C(T) \left[1-\left(\frac{T}{T_c}\right)^2\right]^{3/2}
\end{equation}
where the correction $C(T)$ is approximated by \cite{Semenov2015}
\begin{equation}
C(T) = 0.66 \left[3 - \left(\frac{T}{T_c}\right)^5 \right]^{1/2}
\end{equation}
and the prefactor $J_{\text{dep}}(0)$ is given by  
\begin{equation}
J_{\mathrm{dep}}(0) = \frac{4\sqrt{\pi} \, \text{exp}(2\gamma)}
{21 \zeta(3) \sqrt{3}} \frac{\alpha^{2} (k_B T_c)^{3/2}}{e \rho \sqrt{D\hbar}}.
\end{equation}
Here, $\gamma$ is the Euler–Mascheroni constant, $\zeta(3)$ is the Apéry’s constant, $\rho$ is the resistivity and $\alpha$ is the ratio of the energy gap at zero temperature to $k_{\text{B}}$\tc. 
As no reliable literature values for $\alpha$ in MoSi were found, we used the BCS value $\alpha=1.76$ as it was used for WSi \cite{Zhang2016}. 
We expect that this should be a good estimate since both WSi and MoSi show comparable superconducting properties, as they are both amorphous and silicon based. 
In addition, we note that $C(T)$ may differ in MoSi, as it is a disordered amorphous superconductor compared to NbN.

Another key parameter that governs the metrics of SNSPDs is the interfacial thermal boundary conductance $\beta$. 
This parameter describes the coupling between the superconductor and the substrate. 
It can be determined by measuring the return current $I_{\text{ret}}$ at which the nanowire recovers to the superconducting state \cite{Dane2022}
\begin{equation}
    \beta_\mathrm{eff} = \frac{R_{\text{s}} I_{\text{ret}}^2}{w^2 (T_{\text{hs}}^4 - T_{\text{sub}}^4)}.
    \label{eq:beta_eff}
\end{equation}
At the return current the nanowire is in equilibrium between its self-heating state where electrons are constantly heated with a rate of $I^2 R_{\text{s}} / (w^2 d)$ per unit volume and thermal diffusion and heat flow into the substrate. 
In \eqref{eq:beta_eff} $T_{\text{hs}}$ is the hotspot temperature. 
In several earlier analyses of superconducting nanowires, this temperature has been approximated by \tc \cite{Dharmadurai1979} and we adopt the same approximation here. 
Within this approach, $\beta_\mathrm{eff}$ represents an effective thermal conductance, which phenomenologically captures the overall heat removal from the nanowire to the substrate. 
As such, it incorporates all intermediate processes involved in energy relaxation, without distinguishing between individual contributions. 
\\

Multiple studies used the stoichiometry of superconducting materials as a tool to enhance sensitivity.
For NbTiN Zichi \textit{et al.} found a Nb concentration of 0.62 to be optimal for detection of \SI{1550}{\nano \meter} photons \cite{Zichi2019}. 
Similar studies have been performed for WSi \cite{Baek2011, Luskin2023}. 
The superconducting thin films used in this work have a combination of rather high $R_{\text{s}}$ and low \tc. 
Studies on stoichiometry variation in MoSi thin films indeed showed significant influence on the superconducting properties \cite{Erbe2024, Grotowski2025, Banerjee2017}.
However, MoSi SNSPDs are often only fabricated from thin films based on higher Mo concentrations above 0.68 \cite{Verma2015, Haussler2020, Lita2021}. 
These concentrations result in \tc values in the range of 6-\SI{8}{\kelvin} which allows for operation at liquid helium temperatures, avoiding more complex and costly cryogenic systems. 
In addition, higher overall applicable bias currents lead to larger pulse height enhancing signal-to-noise ratio. 
Our study on the other hand cover a much broader range using thin films having Mo concentrations of 53\% (film~A), 61\% (film~B) and 74\% (film~C). 

\begin{figure*}[t]
    \centering
    \includegraphics[width=\textwidth]{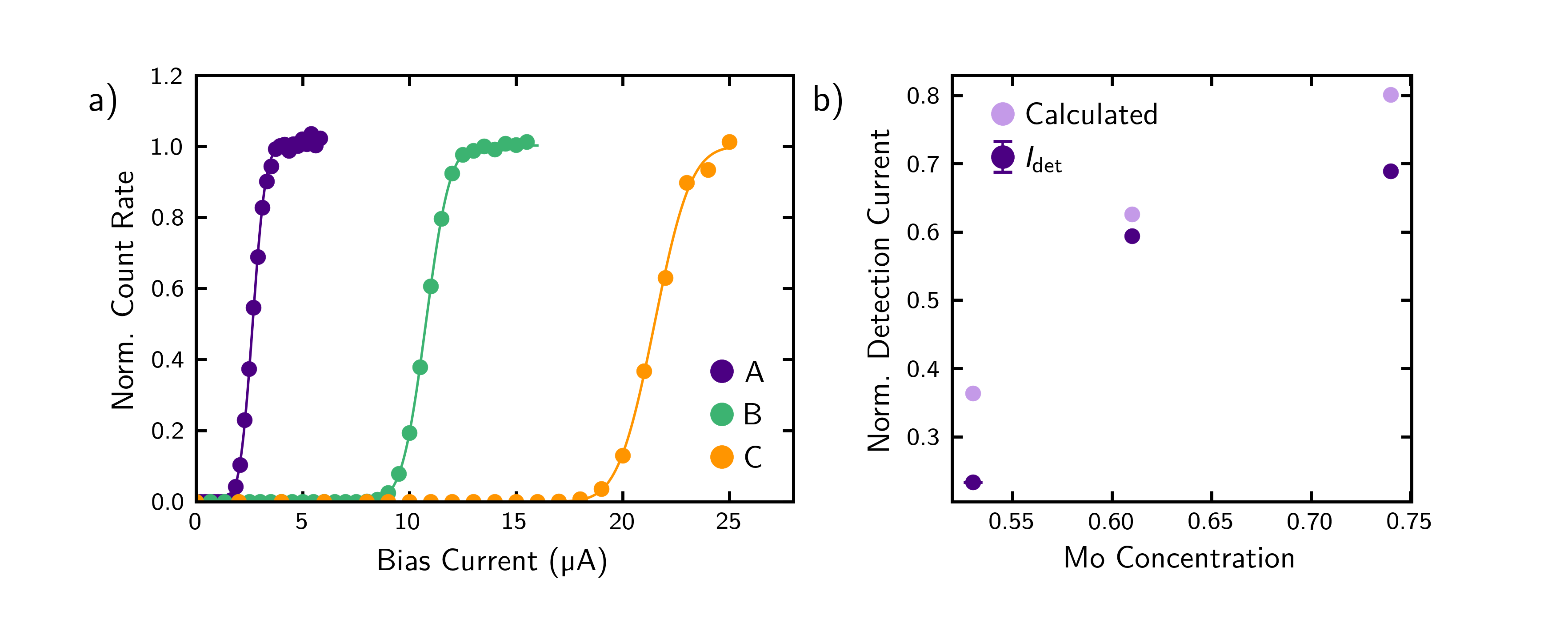}
    \caption{a) Count rate versus bias current curves for \SI{75}{\nano \meter} wire width SNSPDs with different stoichiometries when illuminated at \SI{1550}{\nano \meter} and measured at a temperature of \SI{1}{\kelvin}. b) Normalized detection current for a given wavelength of \SI{1550}{\nano \meter} and comparison with the calculation according to Equation~\eqref{eq:detection_current} using $\xi = 0.15$ and $\tau_{\text{th}} = \SI{7}{\pico \second}$.}
    \label{fig:stoich_CR}
\end{figure*}

We now discuss the SNSPD performance for these three devices. 
The patterned nanowire geometry is similar and consists of a hairpin structure with a nominal nanowire width of \SI{75}{\nano \meter} and a length of \SI{100}{\micro \meter} (see Supp. Mat. \ref{sec:detector_design}). 
We illuminate the devices using a \SI{1550}{\nano \meter} CW laser and perform the measurements at a base temperature of \SI{1}{\kelvin}. 
Figure~\ref{fig:stoich_CR}~(a) presents the count rate versus bias current curves. 
The data is fitted with sigmoidal fits according to Autebert \textit{et al.}~\cite{Autebert2020}
\begin{equation}
    \eta = \frac{\eta_{\text{max}}}{2} \left[ 1 + \text{erf} \left( \frac{I - I_0}{\Delta I} \right) \right].
\end{equation}
The saturation value $\eta_{\text{max}}$ describes the count rate at maximum efficiency, $I_0$ and $I$ are determined by the fit. 
We use $\eta_{\text{max}}$ to normalize the count rate curves. 
Firstly, we observe that the bias current leading to a significant count rate varies a lot for these devices. 
The low \tc device fabricated from film~A with a \mosi{0.53}{0.47} stoichiometry reaches a maximum bias current of only \SI{7}{\micro\ampere}, whereas for the high \tc device, fabricated from film~C, we apply bias levels up to \SI{26}{\micro\ampere}. 
This observation is entirely expected as \idep increases for higher \tc. 
The device fabricated from film~A saturates already at a bias current of \SI{4}{\micro\ampere} whereas the device fabricated from film~C barely saturates despite its higher bias current. 
The device fabricated from film~B lies in between the measurements obtained for films~A and C. 
In order to quantitatively compare the data obtained from the three films we extract the detection current $I_{\text{det}}$, defined as the current where the count rate reaches 50\% of $\eta_{\text{max}}$. 
This point is less sensitive to slight variations in the count rate curves, e.g. due to different settings of the counter threshold levels compared to other points, such as the onset point where first detection event occur. 
Moreover, this point is insensitive to the change of the transition width on the count rate curve with photon energy \cite{Caloz2017}. 
In Figure~\ref{fig:stoich_CR}~(b) we present \idet, each normalized to the depairing current $I_{\text{dep}}$ and the calculated threshold bias at which we expect to observe photon detection given by Equation~\eqref{eq:detection_current}. 
In literature the typical fraction of incident photon energy deposited in the nanowire is in the range $\xi = 0.10-0.17$ \cite{Semenov2005, Lusche2014, Engel2015}. 
We use constant values of $\xi = 0.15$ and $\tau_{\text{th}} = \SI{7}{\pico \second}$ \cite{Ilin2000}. 
For all SNSPDs the measured value of \idet is lower than theoretically estimated values, meaning that lower bias currents are sufficient to register photon detection events. 
A lower normalized detection current enables the detector to be biased further below the critical current while maintaining sensitivity to photons at the same wavelength. 
Although the device fabricated from film~C shows a detection current for \SI{1550}{\nano \meter}, the cutoff wavelength decreases strongly with only small reduction of the bias current. 
This leads to the conclusion that film~A has the highest sensitivity.
An exemplary calculation of the cutoff wavelength in dependency of the normalized bias current for the film parameters is presented in Supp. Mat. \ref{sec:cutoff_calc}.
In comparison of \idet for different Mo concentrations, we see that the offset from the calculated detection current is not constant, indicating a variation of either $\xi$ or $\tau_{\text{th}}$, or both parameters with stoichiometry. \\

\begin{figure}[t]
    \centering
    \includegraphics[width=\columnwidth]{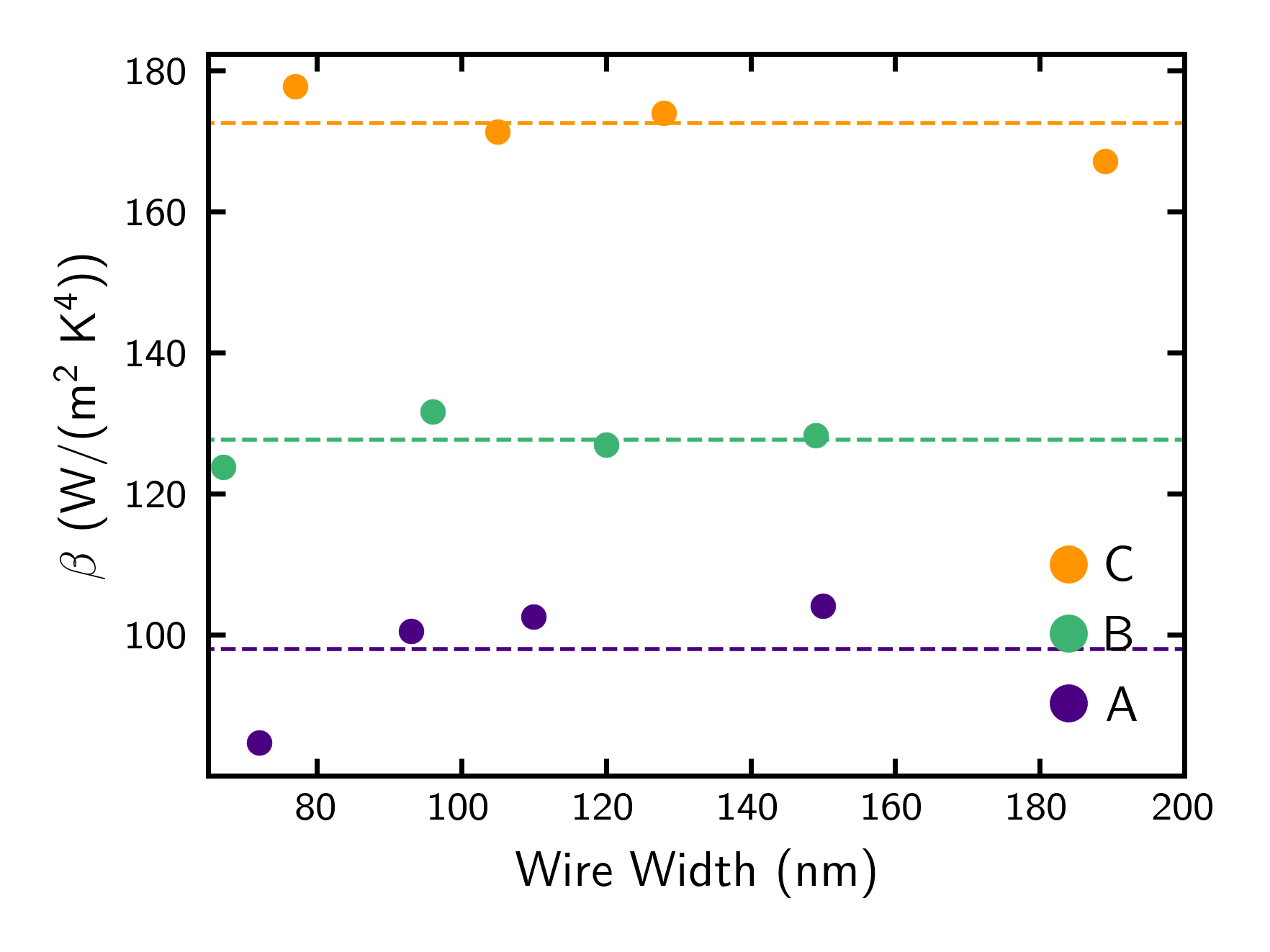}
    \caption{Thermal boundary conductance $\beta_\mathrm{eff}$ extracted from the return current \iret measured at \SI{1}{\kelvin}. The nominal widths are corrected by an effective reduction $\delta w = $ \SI{6.5}{\nano \meter}}
    \label{fig:beta}
\end{figure}

Besides \tc and \rs which change with stoichiometry, our data suggests additional changes in the material parameters which influence the detector performance. 
We probe the thermal coupling between the superconducting thin film and the substrate by measuring the return current and extracting the thermal boundary conductance~$\beta_\mathrm{eff}$. 
As $\beta_\mathrm{eff}$ is defined per area, we expect the thin film interface to remain unchanged for all devices fabricated from the same thin film. 
The initial calculation shows an increase of $\beta_\mathrm{eff}$ with increasing nanowire width, however is should not depend on strip width \cite{Charaev2017a}. 
We conclude that we need to take an additional reduction of the nanowire width originating from oxidation from the sides into account. 
We determine the effective width \sub{w}{eff} by finding a common $\delta w$ for which $\beta_\mathrm{eff}$ is constant. 
Our fitting yields \SI{6.5}{\nano \meter} per side, which is a reasonable value. 
The results of this analysis are presented in Figure~\ref{fig:beta} for different Mo concentrations. 
They show an increase from 98.0 over 127.7 to \SI{173.5}{\watt \per \meter^2 \kelvin^4} with increasing Mo concentration. 
The measured results can be compared with theoretical estimates. 
However, $\beta_\mathrm{eff}$ extracted from electrical measurements represents an effective quantity, which includes all energy relaxation processes within the nanowire. 
In contrast, the following analysis aims to estimate the contribution arising solely from phonon escape across the film–substrate interface. 
As a result, the two quantities are not expected to be equal, and the experimentally extracted values should be regarded as a lumped parameter of the full heat-removal chain rather than a pure interface conductance.
To derive stoichiometry dependent theoretical values of $\beta$, we therefore consider models describing phonon transmission at the interface.
We use the elastic stiffness constants $C_{11}$ and $C_{44}$ of Mo$_{\mathrm{1-x}}$Si$_{\mathrm{x}}$ measured by Djemia \textit{et al.} \cite{Djemia2011}. 
From these values, the corresponding transverse and longitudinal sound velocities $c_\mathrm{t}$ and $c_\mathrm{l}$ are calculated (see Supp. Mat.~\ref{sec:velocities}). 
Within the Debye model and assuming that non-equilibrium phonons escape from the film into the substrate over a characteristic timescale $\tau_{\mathrm{esc}}$, the phonon–substrate heat flow in its  phenomenological form is given by:
\begin{equation}
    \begin{split}
        P_{\mathrm{ph-sub}} =\Sigma_\mathrm{ph-sub}\left(T_\mathrm{ph}^m-T_\mathrm{sub}^m\right),
    \end{split}
\end{equation}
where within the framework of the acoustic mismatch model (AMM) or diffuse mismatch model (DMM), the phonon coupling exponent is $m=4$. $\Sigma_\mathrm{ph-sub}$ is the volumetric phonon-substrate coupling coefficient 
\begin{equation}
    \Sigma_\mathrm{ph-sub} = \beta / d= \frac{\,B_\mathrm{ph}}{4\tau_\mathrm{esc}},
    \label{eq:beta_esc}
\end{equation}
where $B_\mathrm{ph} = \frac{2\pi^2k_\mathrm{B}^4}{5\hbar^3v_\mathrm{avg}^3}$ describes the phonon heat capacity in the Debye limit with the average sound velocity defined as $\frac{1}{v_\mathrm{avg}^3}=\frac{1}{3} \left( \frac{1}{c_\mathrm{l}^3} + \frac{2}{c_\mathrm{t}^3} \right)$. 
Within the ballistic limit $\tau_\mathrm{esc}$ can be written as:
\begin{equation}
\tau_{\mathrm{esc}}=\frac{4d}{u_s\,\bar{\alpha}},
\end{equation}
where $d$ is the film thickness, $u_s = \frac{c_\mathrm{l}^{-2}+2c_\mathrm{t}^{-2}}{c_\mathrm{l}^{-3}+2c_\mathrm{t}^{-3}}$ is the Debye transport-weighted velocity in the film, and $\bar{\alpha}$ is the angle-averaged transmission coefficient of the film-substrate interface. 
Here, the factor $4d$ arises from angular averaging of an isotropic phonon distribution in the film.
\\

\begin{figure}[h]
    \centering
    \includegraphics[width=\columnwidth]{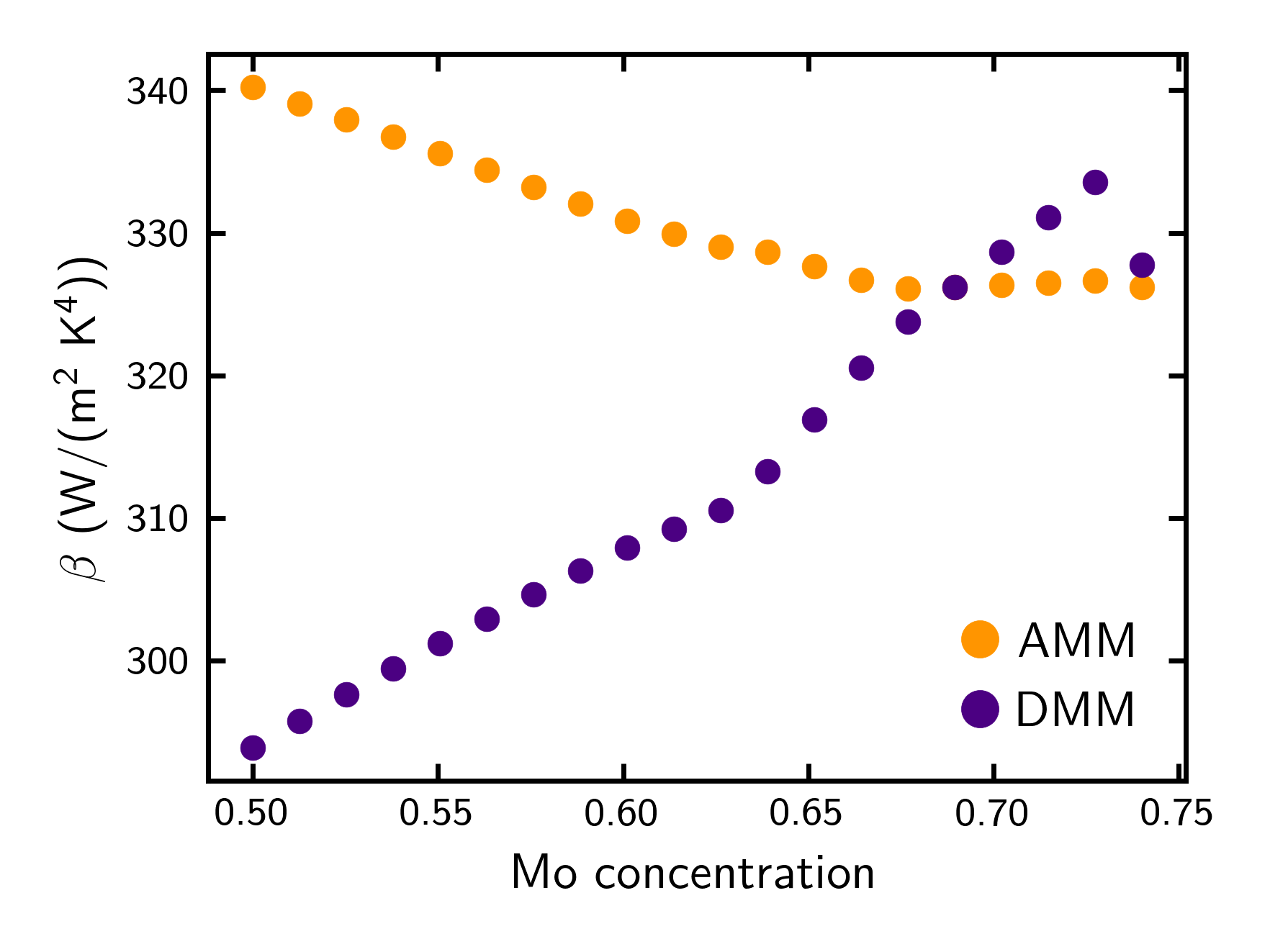}
    \caption{Comparison of thermal boundary conductance values based on AMM and DMM. }
    \label{fig:amm_dmm}
\end{figure}

Because MoSi is amorphous and lacks long-range periodic order, its vibrational modes do not possess a well-defined crystal wavevector. 
As a result, wavevector-conserving approaches such as the AMM are not applicable, making the DMM the more appropriate effective description.
Instead, phonon transmission is evaluated within the DMM, which assumes complete randomization of phonon momentum and polarization at the interface. 
In this framework, the transmission probability is determined by the relative phonon phase space available on each side of the interface.
Within the Debye approximation, the angle-averaged phonon transmission coefficient becomes frequency independent and can be expressed in terms of the longitudinal and transverse sound velocities as:

\begin{equation}
\bar{\alpha}_{\mathrm{DMM}}
=
\frac{c_{\mathrm{sub,l}}^{-2}+2c_{\mathrm{sub,t}}^{-2}}
{c_{\mathrm{MoSi,l}}^{-2}+2c_{\mathrm{MoSi,t}}^{-2}
+
c_{\mathrm{sub,l}}^{-2}+2c_{\mathrm{sub,t}}^{-2}}
\end{equation}
For the SiO$_2$ substrate, we use the sound velocities from Kaplan \cite{Kaplan1979} $c_{\mathrm{SiO_2,l}}=6.09$ km/s and $c_{\mathrm{Sio_2,t}}=4.09$ km/s.
We show the resulting $\beta$ values in Figure~\ref{fig:amm_dmm}. 
We can see that the Mo concentration has a significant influence and shows opposite trends for the AMM and DMM. 
The increase of $\beta$ in the DMM is in good agreement with our experiment, justifying that our system is better described by the DMM. 
The experimentally extracted values are systematically lower than the theoretical predictions, which is expected since the latter describe only the phonon escape contribution, while the experimentally obtained values correspond to an effective thermal conductance including additional energy relaxation processes. \\

\begin{figure*}[t]
    \centering
    \includegraphics[width=\textwidth]{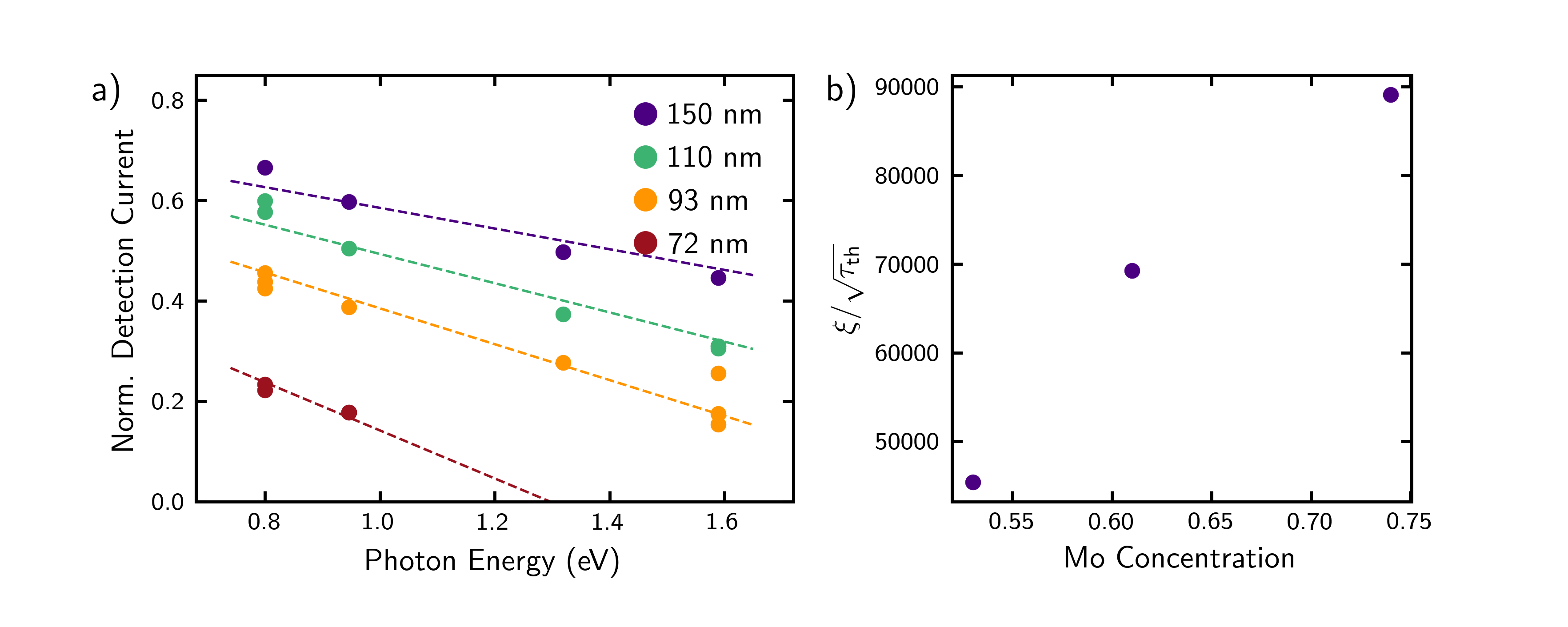}
    \caption{a) Detection current for different photon energies for SNSPDs made from film A. Errorbars are smaller than the size of data points and therefore not displayed. b) Fitted $\xi$ values for film A and B. c) Thermal boundary conductance $\beta$ calculated from measurements of the return current of film A.}
    \label{fig:wirewidth}
\end{figure*}

Previous studies of other superconducting materials revealed a significant influence of the wire width on the device performance \cite{Chang2022, Schuck2013}. 
To proceed with our analysis we investigate different wire widths from \SI{67}{\nano \meter} to \SI{189}{\nano \meter} fabricated from the thin films. 
For each wire width we measure count rate curves under illumination with wavelengths of \SI{780}{\nm}, \SI{940}{\nano \meter}, \SI{1310}{\nano \meter} and \SI{1550}{\nano \meter}, respectively. 
The extracted detection currents $I_{\text{det}}$ are plotted in Figure~\ref{fig:wirewidth}~(a) as a function of photon energy. 
The detection current for higher photon energies for the smallest nanowire width is measured, but excluded in the fitting procedure, as the detection occurs at a bias current, where voltage pulses are within the noise level. 
We observe a clear decrease of the detection current with nanowire width, as predicted by the hotspot diffusion model. 
Thus, we expect both the conversion efficiency $\xi$ and $\tau_{\text{th}}$ to be stoichiometry dependent, but independent on wire width.
In the present work, however, we assume $\tau_\mathrm{th}$ to be constant in order to reduce the number of free parameters and enable a direct comparison between devices. 
We reuse Equation~\eqref{eq:detection_current} to fit our data for the nanowire width variation with $\xi' = \xi/\sqrt{\tau_{\text{th}}}$ as a combined fit parameter. 
As a result, variations observed in the fitted parameter $\xi'$ should be interpreted as an effective combination of changes in both parameters. 
A quantitative separation of these contributions would require independent knowledge of the relevant electron–phonon interaction parameters, which is beyond the scope of this study.
We take the previously determined reduction of the nominal nanowire width from oxidization into account, as well as the diffusivity of our thin films, which is $D = \SI{0.45}{\centi \meter^2 \per \s}$. 
This way, we confirm the expected linear dependency of the cutoff bias, as demonstrated previously for WSi \cite{Gaudio2016} and TaN \cite{Lusche2014}. 
Linear extrapolation to zero cutoff reveals an intersection point between 0.62 and 0.92 of the normalized detection current. 
This indicates, that the critical current which can be applied to the nanowire to reach ultimate detection is not the depairing current, but limited by other factors, such as side wall defects from nanofabrication. 
Similar discrepancy was observed in previous studies \cite{Lusche2014}. \\

The resulting $\xi'$ parameters from the fit of all three stoichiometries are shown in Figure~\ref{fig:wirewidth}~(b). 
We see an increase with increasing Mo concentration. 
Since $\xi' = \xi/\sqrt{\tau_{\text{th}}}$, this trend reflects either an increase in the photon energy conversion efficiency $\xi$, a decrease in $\tau_\mathrm{th}$, or a combination of both. 
Assuming $\tau_\mathrm{th}$ to be approximately constant, this would imply that a larger fraction of the absorbed photon energy contributes to hotspot formation at higher Mo concentrations. 
Considering $\beta_\mathrm{eff}$ alone, the opposite trend would be expected. 
A lower $\beta_\mathrm{eff}$ leads to reduced heat leakage into the substrate, thereby favoring energy confinement in the nanowire and enhancing detection efficiency. 
It has been shown that a reduced thermal coupling leads to an increased hotspot relaxation time \cite{Xu2023} and favors not only detection efficiency \cite{Ota2013}, but also enhances sensitivity for longer wavelength photons \cite{Yin2024}.
Based on this argument, one would anticipate higher conversion efficiency for lower Mo concentration, in contrast to our observations. 
This indicates that the detection efficiency is not governed solely by heat removal to the substrate, but is strongly influenced by internal energy relaxation processes.

Other properties to investigate are the relevant interaction timescales, especially of electron-phonon interaction. 
The equilibrium relation states $C_{\text{e}}/C_{\text{ph}} = \tau_{\text{e-ph}}/\tau_{\text{ph-e}}$. 
Vodolazov \cite{Vodolazov2017} showed that with larger specific heat ratio $C_{\text{e}}/C_{\text{ph}}$ ratios, for the same amount of energy deposited in the nanowire a larger fraction photon energy goes into the electronic system and less into the phononic bath.

At the same time, estimates of the phonon heat capacity coefficient $B_\mathrm{ph}$ within the Debye approximation suggest an increase with Mo concentration. 
However, this model provides only a simplified description of the phonon spectrum. 
In ultrathin amorphous MoSi films, disorder-induced modifications of the vibrational density of states can lead to significant deviations from Debye behavior. 
In this context, it has been reported that the phonon heat capacity relevant for hotspot dynamics can be substantially lower than the value expected for three-dimensional Debye phonons, indicating a reduced phonon contribution to energy relaxation \cite{Sidorova2018}. 
The correlated driving parameter for the hotspot growth is hence $\tau_{\text{ph-e}}$. 
It is estimated to be between \SI{100}{\pico \second} to \SI{200}{\pico \second} for thin WSi films \cite{Sidorova2018} compared to only few tens of ps for NbN and NbTiN \cite{Sidorova2021}. 
Hence also the ratios differ, as for NbN  \sub{\tau}{e-ph}/\sub{\tau}{e-e} $\approx 1$ \cite{Korneeva2017} compared to WSi \sub{\tau}{e-ph}/\sub{\tau}{e-e} $\approx 3.8$ \cite{Zhang2018a}. 
This way the photon energy is deposited into in the electron subsystem and remains there for a longer time, leading to enhanced growth of the hotspot. 
More generally, higher conversion efficiencies are observed for WSi compared to NbN.
We can compare our results with the data obtained from Zhang \textit{et al.} \cite{Zhang2018a} for WSi as we expect our results to behave in a similar way, due to their similar amorphous character. 
For our results we speculate that we not only have high enhancement of the sensitivity when comparing different material platforms, but also when increasing the Mo concentration and changing the stoichiometry. 
The exact electron-phonon interaction times for calculations of the specific heat capacities can be extracted from magnetoconductance measurements. 
Besides stoichiometry, another method to influence $C_{\text{ph}}$ is to reduce device dimensionality (i.e. nanowire thickness) because due to confinement of the phonon modes the number of allowed phonon wavevectors is reduced \cite{Sidorova2023}. 
Consequently, with reduced film thickness the sensitivity is expected to increase \cite{Sidorova2020, Semenov2009}. 
\\

Sensitive, low \tc SNSPDs usually come with the downside, that they have to be operated at very low temperatures below \SI{1}{\kelvin}. 
We study the count rate behavior of the detector for our low \tc device fabricated from film~A for different operation temperatures and we can confirm reaching unitary internal quantum efficiency of our devices to bath temperatures up to \SI{2.5}{\kelvin} (see Supp. Mat.~\ref{sec:temp_dependency}).
High \tc SNSPDs have to be biased very close to the critical current to show sufficient sensitivity, where they have an increased dark count rate. 
On the other hand, higher absolute bias currents lead to larger output pulses, improving the signal-to-noise ratio and improved timing jitter. 
However, a low signal-to-noise ratio for low absolute bias currents can be overcome by using cryogenic amplification in short distance to the SNSPD. 
As in most cases, a trade off exists between enhanced sensitivity, especially towards longer wavelengths and other performance metrics. 
Our study allows to predict the sensitivity of SNSPDs, taking the interfacial thermal boundary conductance into account, which can be used for further thin film optimizations. 

\section{Summary and Conclusion}
\vspace{-10pt}
In summary, we analyzed the transport properties of MoSi thin films, utilizing a universal scaling law and obtained the fitting parameters $A = 20214$ and $B = 1.13$, which is in good agreement with previous values found in literature. 
By comparing the sensitivity for SNSPDs of three different stoichiometries, the device fabricated from \mosi{0.53}{0.47} was found to have highest sensitivity measured with \SI{1550}{\nano \meter} photons. 
Measurements of the return current show as well a decrease of $\beta_\mathrm{eff}$ for lower Mo concentrations. 
The study of the nanowire width dependency confirms the linear dependence of the detection current $I_{\text{det}}$ with photon energy. 
Furthermore, we extracted the combined parameter $\xi' = \xi / \sqrt{\tau_\mathrm{th}}$, which is increasing with Mo concentration.
From this we conclude, that the photon energy conversion ratio is increased with higher Mo concentration. 
We compare the relevant time scales for this process with other material platforms.
Finally, we presented the temperature dependence of our most sensitive \mosi{0.53}{0.47} device, showing that the device can be operated efficiently at temperatures up to \SI{2.5}{\kelvin}. \\

We gratefully acknowledge support from the German Federal Ministry of Research Technology and Space (BMFTR) via the funding program “Quantum technologies – from basic research to market” (projects PhotonQ (13N15760), QPIS.2 (16KISQ172), FgED (13N17529), QPIS (16K1SQ033), QPIC-1 (13N15855), SPINNING (13N16214) and QR-N (16KIS2197)), as well as from the German Research Foundation (DFG) under Germany’s Excellence Strategy EXC-2111 (390814868) and projects INST 95/1720-1 (MQCL) and PQET (INST 95/1654-1). This research is also supported via of the ‘Munich Quantum Valley”, which is supported by the Bavarian state government with funds from the “Hightech Agenda Bayern Plus”.

\section{Supplementary Material}
\vspace{-10pt}

\subsection{Detector Design}
\vspace{-10pt}
\label{sec:detector_design}

\begin{figure}[h]
    \centering
    \includegraphics[width=\columnwidth]{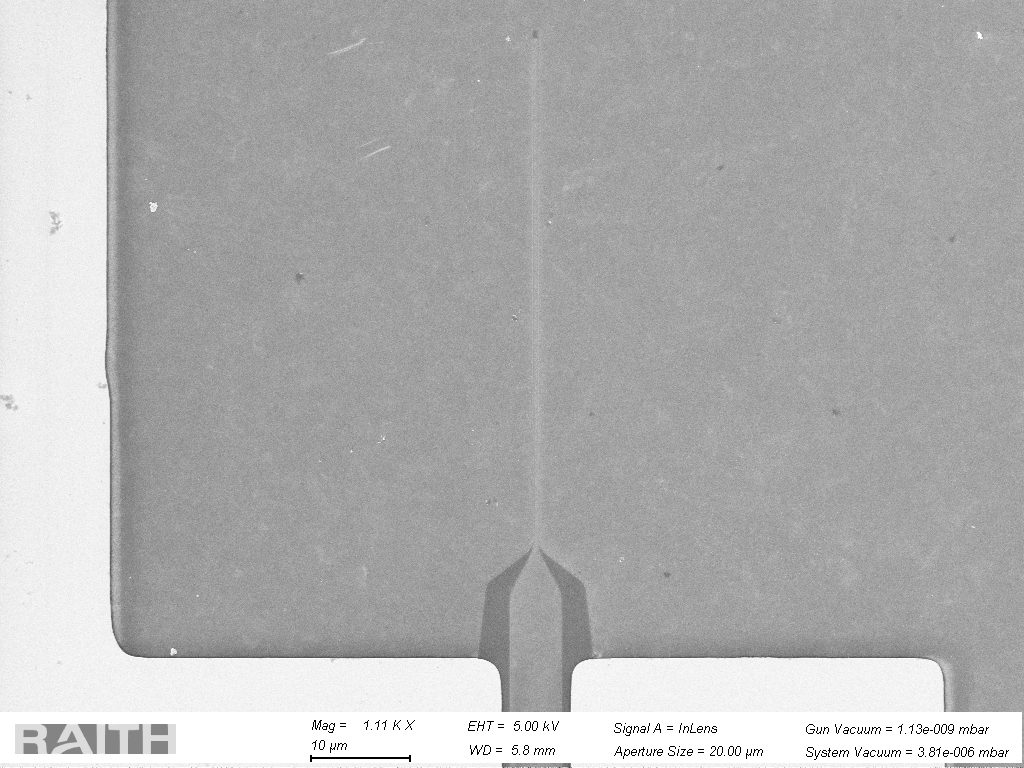}
    \caption{SEM image of one example hairpin structure.}
    \label{fig:hairpin}
\end{figure}

The detectors measured have a hairpin geometry. The nanowire is a single meander with a total length of \SI{100}{\micro \meter}. The gap between the meanders is kept constant at \SI{200}{\nano \meter}, whereas the nanowire width is varied from \SI{67}{\nano \meter} to \SI{190}{\nano \meter}. 
Figure~\ref{fig:hairpin} shows an SEM image of one  detector. 

\subsection{Calculation of Cutoff Wavelength}
\vspace{-10pt}
\label{sec:cutoff_calc}

\begin{figure}[h]
    \centering
    \includegraphics[width=\columnwidth]{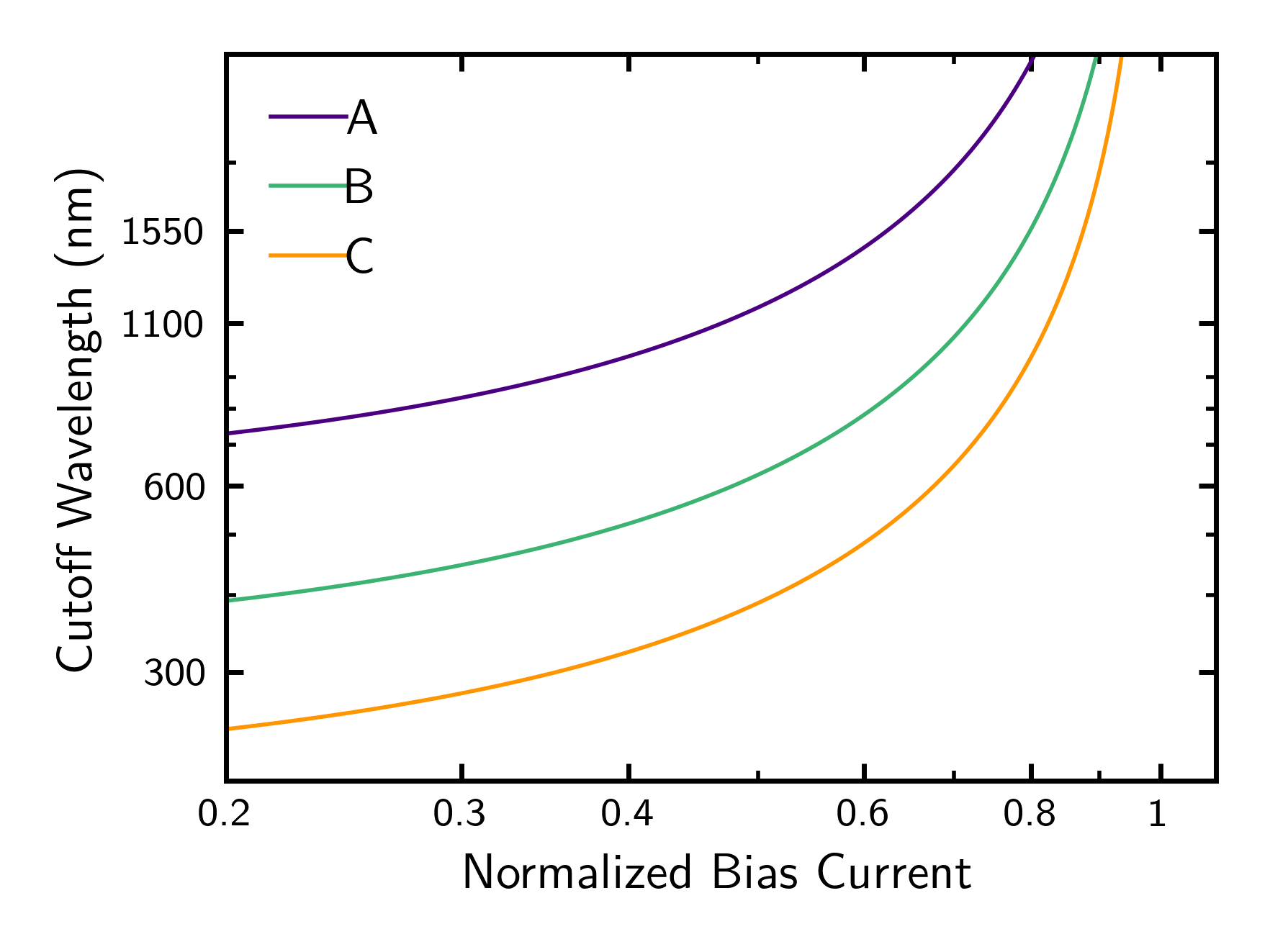}
    \caption{Calculation of cutoff wavelength for different bias currents normalized to the depairing current at $\xi = 0.15$ and $\tau_{\text{th}} = \SI{7}{\pico \second}$.}
    \label{fig:cutoff_calc}
\end{figure}

Equation~\eqref{eq:cutoff} suggests that the cutoff wavelength diverges for bias currents close to the depairing current. 
The calculated cutoff wavelength for fixed values of $\xi = 0.15$ and $\tau_{\text{th}} = \SI{7}{\pico \second}$ is shown in Figure~\ref{fig:cutoff_calc}. 
This raises the assumption that every detector should be able to detect similar wavelengths with very low photon energies. 
However, in the experiment the cutoff wavelength is limited by the switching current of the device, which is in reality lower than the theoretical depairing current. 
Usually, quality factors of maximum 0.6-0.7 have been reported in literature. 

\subsection{Transverse and Longitudinal Sound Velocities of MoSi}
\begin{figure}[h]
    \centering
    \includegraphics[width=\linewidth]{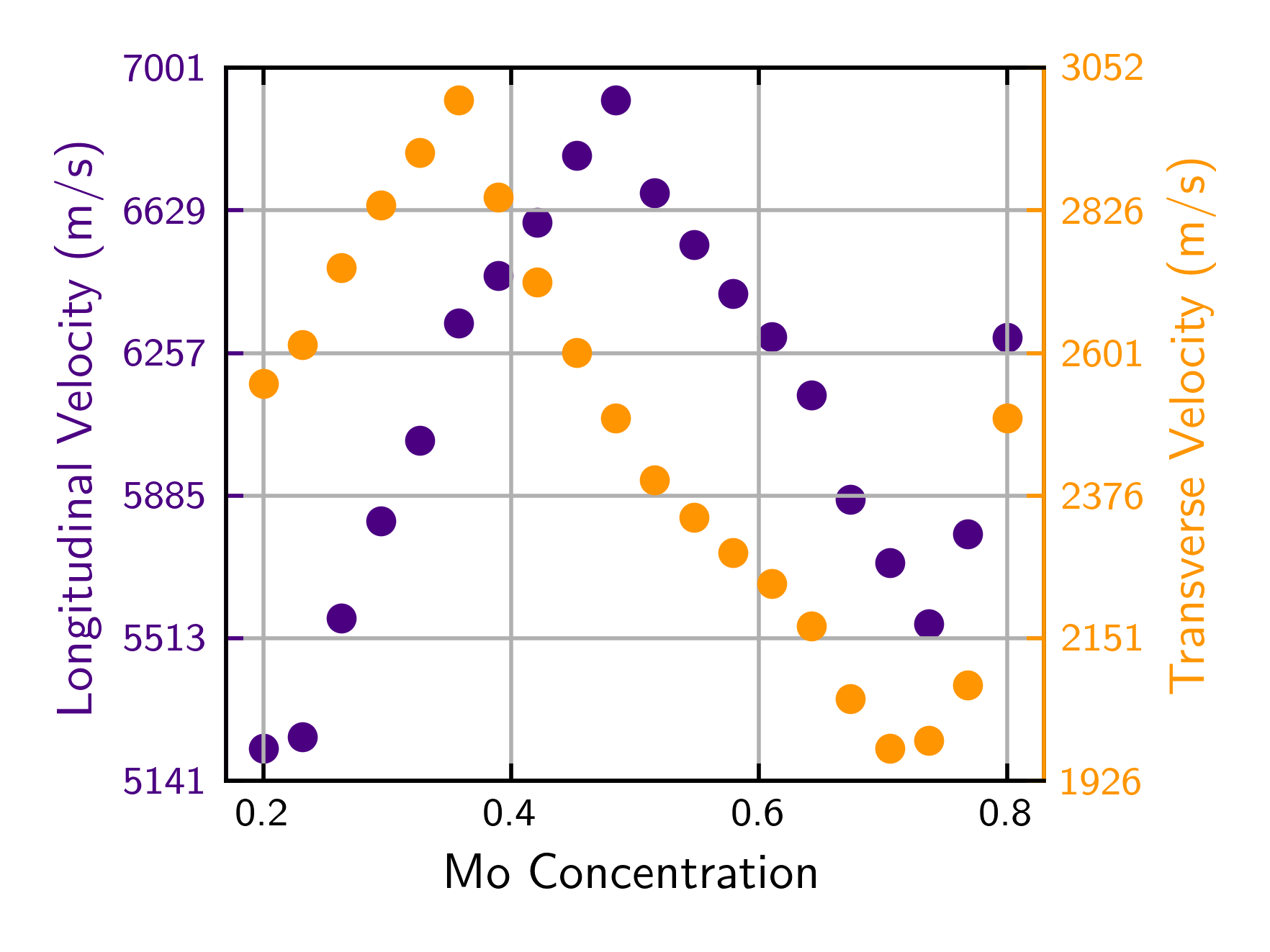}
    \caption{Stoichiometry dependence of the longitudinal and transverse sound velocities in Mo$_{\mathrm{1-x}}$Si$_{\mathrm{x}}$.}
    \label{fig:velocities}
\end{figure}

\vspace{-10pt}
\label{sec:velocities}
In crystalline solids, the velocity of acoustic waves generally depends on the direction of propagation with respect to the crystallographic axes, as the elastic response is anisotropic. In contrast, amorphous materials lack long-range order and can therefore be treated, to a good approximation, as isotropic elastic media. 
For an isotropic solid, the elastic stiffness tensor can be expressed using the Lamé coefficients $\lambda$ and $\mu$ as:
\begin{equation}
C_{11} = \lambda + 2\mu, \qquad
C_{12} = \lambda, \qquad
C_{44} = \mu 
\end{equation} 
The propagation of elastic waves in an isotropic medium gives rise to two acoustic modes: a longitudinal mode and a transverse mode. 
Their velocities are determined by the elastic constants and the mass density $\rho$. 
The corresponding sound velocities are:
\begin{equation}
c_l = \sqrt{\frac{\lambda + 2\mu}{\rho}}, \qquad
c_t = \sqrt{\frac{\mu}{\rho}} 
\end{equation}
Using the relations between the Lamé coefficients and the elastic stiffness constants, it follows that these expressions can be written directly in terms of $C_{11}$ and $C_{44}$ \cite{Mason1956} as

\begin{equation}
c_l = \sqrt{\frac{C_{11}}{\rho}}, \qquad
c_t = \sqrt{\frac{C_{44}}{\rho}}. 
\end{equation}
This formulation allows the longitudinal and transverse sound velocities of amorphous Mo$_{\mathrm{1-x}}$Si$_{\mathrm{x}}$ to be determined from the experimentally measured elastic stiffness constants reported by Djemia \textit{et al.} \cite{Djemia2011}.
The data obtained from the reference measured film thicknesses of \SI{250}{\nano \meter}, therefore we expect the real values to be about 10\% smaller than in the calculation. 
The results are shown in Figure~\ref{fig:velocities}.

\subsection{Temperature Dependent Sensitivity}
\vspace{-10pt}
\label{sec:temp_dependency}
SNSPDs fabricated from silicides require lower operating temperatures than NbN or NbTiN SNSPDs, due to lower \tc. 
Lower operating temperatures reduce the dark count rate significantly \cite{Yamashita2010}. 
\begin{figure}[h]
    \centering
    \includegraphics[width=\columnwidth]{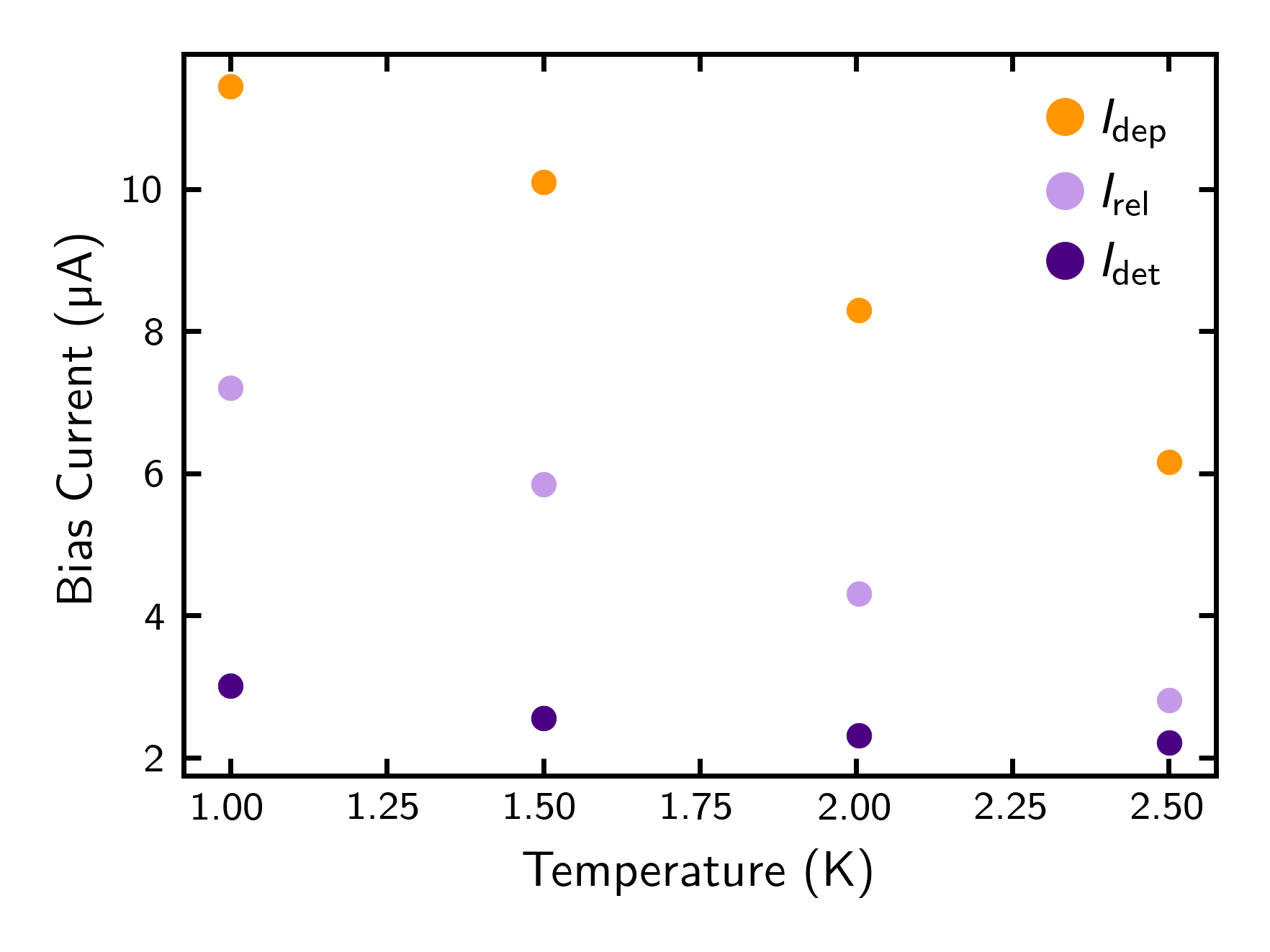}
    \caption{Characteristic currents for toe, detection, saturation and relaxation current for different operational temperature for the same device.}
    \label{fig:temperature_90mA}
\end{figure}
Our SNSPDs fabricated from \mosi{0.53}{0.47} have a \tc of \SI{4.1}{\kelvin} and we tested their performance at different operating temperatures $\leq$ \SI{2.5}{\kelvin}. 
Figure~\ref{fig:temperature_90mA} summarizes the characteristic current values for a device with \SI{72}{\nano \meter} wire width. 
\irel is the maximum applicable bias current, before the SNSPD enters the relaxation oscillations regime \cite{Liu2012}, occurring due to the \SI{25}{\ohm} shunt resistor. 
Although the plateau length decreases with increasing temperature, the device shows saturation at a base temperature of \SI{2.5}{\kelvin}.

\bibliography{references.bib}

\end{document}